\documentclass[journal]{IEEEtran}
\IEEEoverridecommandlockouts
\usepackage[dvips]{graphicx}
\usepackage[fleqn]{amsmath}
\usepackage[varg]{txfonts}
\usepackage{colortbl}
\usepackage{algorithm}
\usepackage{algorithmic}
\usepackage{tabularx}
\usepackage{cite}
\usepackage{graphicx}

\ifCLASSINFOpdf
\else
\fi
\begin{document}
%\title{Bare Demo of IEEEtran.cls for Journals}
\title{A Game-Theoretic Approach to Energy-Efficient Resource Allocation in Device-to-Device Underlay Communications}

%\author{\IEEEauthorblockN{Zhenyu Zhou$^1$, Mianxiong Dong$^2$, Kaoru Ota$^3$, Ruifeng Shi$^1$, Zhiheng Liu$^4$, Takuro Sato$^5$}\\
%\IEEEauthorblockA{$^1$State Key Laboratory of Alternate Electrical Power System with Renewable Energy Sources,\\
%School of Electrical and Electronic Engineering, North China Electric Power University, Beijing, China, 102206\\
%Email: zhenyu\_zhou@fuji.waseda.jp\\
%$^2$National Institute of Information and Communications Technology, Kyoto, Japan\\
%$^3$Department of Information and Electric Engineering, Muroran Institute of Technology, Muroran, Hokkaido, Japan\\
%$^4$Department of Network Technology, China Mobile Communication Corporation Research Institute, Beijing, China\\
%$^5$Graduate School of Global Information and Telecommunication Studies, Waseda University, Tokyo, Japan}}

\author{Zhenyu~Zhou, Mianxiong Dong, Kaoru Ota, Ruifeng Shi, Zhiheng Liu, and Takuro Sato
\thanks{Zhenyu Zhou and Ruifeng Shi are with the State Key Laboratory of Alternate Electrical Power System with Renewable Energy Sources, School of Electrical and Electronic Engineering, North China Electric Power University, Beijing, China, 102206 (Email: zhenyu\_zhou@fuji.waseda.jp). }
\thanks{Mianxiong Dong is with the National Institute of Information and Communications Technology, Kyoto, Japan (Email: mxdong@ieee.org). }
\thanks{Kaoru Ota is with the Department of Information and Electric Engineering, Muroran Institute of Technology, Muroran, Hokkaido, Japan.}
\thanks{Zhiheng Liu is with the Department of Network Technology, China Mobile Communication Corporation Research Institute, Beijing, China.}
\thanks{Takuro Sato is with the Graduate School of Global Information and Telecommunication Studies, Waseda University, Tokyo, Japan}
}
    %    John~Doe,~\IEEEmembership{Fellow,~OSA,}
        %and~Jane~Doe,~\IEEEmembership{Life~Fellow,~IEEE}% <-this % stops a space
%\thanks{M. Shell is with the Department
%of Electrical and Computer Engineering, Georgia Institute of Technology, Atlanta,
%\thanks{Manuscript received April 19, 2005; revised January 11, 2007.}}


\markboth{IET Communications,~Vol.~XX, No.~XX, XX~XX}%
{Shell \MakeLowercase{\textit{et al.}}: Bare Demo of IEEEtran.cls for Journals}



\maketitle

\begin{abstract}
%\boldmath
Despite the numerous benefits brought by Device-to-Device (D2D) communications, the introduction of D2D into cellular networks poses many new challenges in the resource allocation design due to the co-channel interference caused by spectrum reuse and limited battery life of User Equipments (UEs). Most of the previous studies mainly focus on how to maximize the Spectral Efficiency (SE) and ignore the energy consumption of UEs. In this paper, we study how to maximize each UE's Energy Efficiency (EE) in an interference-limited environment subject to its specific Quality of Service (QoS) and maximum transmission power constraints. We model the resource allocation problem as a noncooperative game, in which each player is self-interested and wants to maximize its own EE. A distributed interference-aware energy-efficient resource allocation algorithm is proposed by exploiting the properties of the nonlinear fractional programming. We prove that the optimum solution obtained by the proposed algorithm is the Nash equilibrium of the noncooperative game. We also analyze the tradeoff between EE and SE and derive closed-form expressions for EE and SE gaps. 
\end{abstract}


% \fi
\IEEEpeerreviewmaketitle

\section{Introduction}
Device-to-Device (D2D) communications allows two User Equipments (UEs) that are in the proximity of each other to exchange information over a direct link, and can be operated as an underlay to cellular networks by reusing the scarce spectrum resources \cite{D2D_LTE}. As a result, D2D communications underlaying cellular networks bring numerous benefits including the proximity gain, the reuse gain, and the hop gain \cite{D2D_design}. The applications and research challenges of D2D communications for current and future cellular networks were studied in \cite{D2D_5G, D2D_loadbalance2014}, and the corresponding standardization activities in Third Generation Partnership Project (3GPP) were introduced in \cite{D2D_3GPP}.

However, despite the numerous benefits brought by D2D communications,the introduction of D2D communications into cellular networks poses many new challenges in the resource allocation design due to the co-channel interference caused by spectrum reuse and limited battery life of UEs. A large number of works have been done on how to perform resource allocation to increase Spectral Efficiency (SE) (or throughput) in an interference-limited environment. A Stackelberg game based resource allocation scheme was proposed in \cite{Zhaoyang_ICC2013}, in which the Base Station (BS) and D2D UEs were modeled as the game leader and followers respectively. Another Stackelberg game based scheme was proposed in \cite{Feiran_WCNC2013}, in which cellular UEs rather than the BS were modeled as game leaders. A two-stage resource allocation scheme which employs both the centralized and distributed approaches was proposed in \cite{Lee_two_stage}. A three-stage resource allocation scheme which combines admission control, power allocation, and link selection was proposed in \cite{D2D_Li}. A reverse Iterative Combinatorial Auction (ICA) based resource allocation scheme was proposed in \cite{Song_JSAC} for optimizing the system sum rate. The resource allocation problems in relay-aided scenarios were studied in \cite{Hasan_2014, intracluster_2013}, and in infeasible systems where all users can not be supported simultaneously were studied in \cite{IET_D2D}. The throughput performance of the D2D underlay communications with different resource sharing modes was evaluated in \cite{Doppler_TWC}. SE enhancement of D2D communications for wireless video networks was studied in \cite{D2D_video2014}. Resource allocation for D2D communications underlaying cellular networks powered by renewable energy sources was studied in \cite{D2D_greencell2014}. A comprehensive overview and discussion of resource management for D2D underlay communications is provided in \cite{Song_Springer2014}.

The above mentioned works mainly focus on how to maximize SE and ignore the energy consumption of UEs. In practical implementation, UEs are typically handheld devices with limited battery life and can quickly run out of battery if the energy consumption is ignored in the system design. Therefore, in this paper, we focus on how to optimize the Energy Efficiency (EE) (defined as bits/Hz/J) through resource allocation in an interference-limited environment. Unfortunately, optimum EE and SE are not always achievable simultaneously and may sometimes even conflict with each other \cite{EE_SE_tradeoff}. The EE and SE tradeoff for D2D communications have been studied in \cite{Zhou_WCL2014, D2D_mobility}.

For the EE maximization problem, distributed resource allocation algorithms which are based on either the reverse ICA game or the bisection method were proposed in \cite{Feiran_2012} and \cite{EE_analysis} respectively. However, the authors have not considered the Quality of Service (QoS) provisioning constraints and have not derived a close-form solution. Centralized resource allocation algorithms for optimizing EE in the Device-to-MultiDevice (D2MD) or D2D-cluster scenarios were studied in \cite{D2D_MIMO} and \cite{Si_cluster} respectively. One major disadvantage of the centralized algorithms is that the computational complexity and signaling overhead increase significantly with the number of UEs.  Besides, since the optimization process is carried out in the BS, the optimum solution needs to be delivered to the UEs within the channel coherence time. Instead of maximizing EE, auction-based resource allocation scheme and D2D cooperative relays were proposed to improve battery lifetime in \cite{Feiran_ICC2013} and \cite{Ta_ICC2014} respectively. Fractional Frequency Reuse (FFR) based two-stage resource allocation algorithm was proposed in \cite{Mumtaz_ICC2014}. %A Stackelberg game based energy-aware resource allocation scheme was proposed in \cite{Zhou_ICCS2014} to optimize the weighted sum of achievable transmission rate, power consumption, and interference revenue (or cost) based on multi-objective optimization (MOO) methods \cite{MOO}. 
Coalition game based resource sharing algorithms were proposed in \cite{Wu_TVT2014, Wu_CL2014} to jointly optimize the model selection and resource scheduling. The authors assumed that independent D2D UEs and cellular UEs can communicate with one another and act together as one entity to improve their EE in the game.

In this paper, firstly, we propose a distributed interference-aware energy-efficient resource allocation algorithm to maximize each UE's EE subject to the QoS provisioning and transmission power constraints. Since either cellular UEs or D2D UEs are selfish and are only interested in maximizing their own individual utility, which may be even conflicting with each other. In order to solve this problem, we adopt a game-theoretic approach to model the resource allocation problem as a noncooperative game in which each player is self-interested and wants to maximize its own EE. Game theory provides a tool set for analyzing optimization problems with multiple conflicting objective functions and has been widely used for resource allocation in D2D communications \cite{game_theory_1994, Feiran_2012, Feiran_ICC2013, Wu_TVT2014, Wu_CL2014, SongMag2014}. Compared to the cooperative game model used in \cite{Wu_TVT2014, Wu_CL2014}, the noncooperative model has the advantage of a lower overhead for information exchange among UEs. Both of the D2D UEs and cellular UEs are taken into consideration. The EE utility function of each player is defined as the SE divided by the total power consumption, which includes both transmission and circuit power. The formulated EE maximization problem is non-convex but can be transformed into a convex optimization problem by using the nonlinear fractional programming developed in \cite{Dinkelbach}. Then we prove that a Nash equilibrium exists in the noncooperative game, and the optimum solution obtained by the proposed algorithm is exactly the Nash equilibrium. We also derive a spectral-efficient algorithm and compare it with the proposed energy-efficient algorithm through computer simulations. Finally, we analyze the tradeoff between EE and SE in an interference-limited environment and derive closed-form expressions of EE and SE gaps for D2D and cellular UEs respectively.

The structure of this paper is organized as follows: Section \ref{System Model} introduces the system model of the D2D communication underlaying cellular networks. Section \ref{distributed} introduces the distributed iterative optimization algorithm for maximizing each UE's EE. Section \ref{distributed_SE} introduces the distributed spectral-efficient resource allocation algorithm for the purpose of comparison. Section \ref{tradeoff} introduces the tradeoff between EE and SE for the energy-efficient and spectral-efficient algorithms. Section \ref{Simulation Results} introduces the simulation parameters, results and analyses. Section \ref{Conclusion} gives the conclusion.

\section{System Model}
\label{System Model}

\begin{figure}[t]
\begin{center}
\scalebox{0.36} 
{\includegraphics{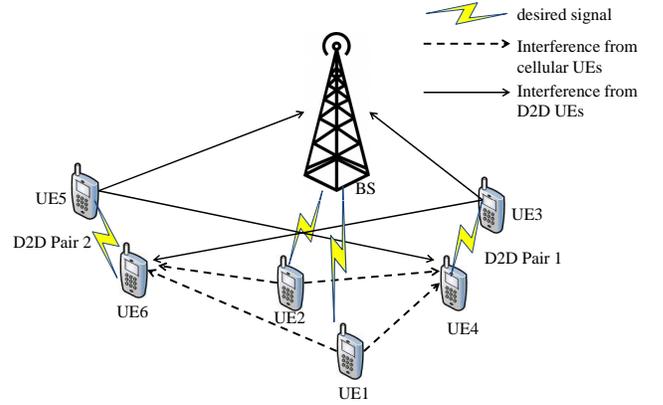}}
\end{center}
\caption{System model of D2D communications with uplink channel reuse.}
\label{D2D_model}
\end{figure}

In this paper, we consider the uplink scenario of a single cellular network, which is composed of the base station, D2D UEs, and cellular UEs. Fig. \ref{D2D_model} shows the system model of D2D communications with uplink resource sharing. There are two cellular UEs ($\mbox{UE}_1$ and $\mbox{UE}_2$), and two D2D pairs ($\mbox{UE}_3$ and $\mbox{UE}_4$, and $\mbox{UE}_5$ and $\mbox{UE}_6$ respectively). A pair of D2D transmitter and receiver forms a D2D link, and a cellular UE and the BS form a cellular link. The UEs in a D2D pair are close enough to enable D2D communications. Each cellular UE is allocated with an orthogonal link (e.g., an orthogonal resource block in LTE), i.e., there is no co-channel interference between cellular UEs. At the same time, the two D2D pairs reuse the same channels allocated to cellular UEs in order to improve SE. As a result, the BS suffers from the interference caused by the D2D transmitters ($\mbox{UE}_3$ and $\mbox{UE}_5$), and the D2D receivers ($\mbox{UE}_4$ and $\mbox{UE}_6$) suffer from the interference caused by cellular UEs ($\mbox{UE}_1$ and $\mbox{UE}_2$) and the other D2D transmitters that reuse the same channel ($\mbox{UE}_5$ or $\mbox{UE}_3$ respectively).

The set of UEs is denoted as $\mathcal{S}=\{ \mathcal{N}, \mathcal{K} \}$, where $\mathcal{N}$ and $\mathcal{K}$ denote the sets of D2D UEs and cellular UEs respectively. The total numbers of D2D links and cellular links are denoted as $N$ and $K$ respectively. The Signal to Interference plus Noise Ratio (SINR) of the $i$-th D2D pair ($i \in \mathcal{N}$) in the $k$-th ($k \in \mathcal{K}$) channel is given by
\begin{align}
\label{eq:SINRD}
\gamma_i^k = \frac{p_i^k g_{i}^k}{p_c^k g_{c, i}^k+\sum_{j=1, j\neq i}^{N}p_{j}^k g_{j, i}^k+N_0}, 
\end{align}
where $p_i^k$, $p_c^k$, and $p_{j}^k$ are the transmission power of the $i$-th D2D transmitter, the $k$-th cellular UE, and the $j$-th D2D transmitter in the $k$-th channel respectively. $g_{i}^k$ is the channel gain of the $i$-th D2D pair, $g^k_{c, i}$ is the interference channel gain between the $k$-th cellular UE and the $i$-th D2D receiver, and $g_{j, i}^k$ is the interference channel gain between the $j$-th D2D transmitter and the $i$-th D2D receiver. $N_0$ is the noise power. $p_c^k g^k_{c, i}$ and $\sum_{j=1, j\neq i}^{N} p_{j}^k g_{j, i}^k$ denote the interference from the cellular UE and the other D2D pairs that reuse the $k$-th channel respectively.

The received SINR of the $k$-th cellular UE at the BS is given by
\begin{align}
\label{eq:SINRC}
\gamma_c^k = \frac{p_c^k g_c^k}{\sum_{i=1}^{N}p_{i}^k g_{i, c}^k+N_0}, 
\end{align}
where $g_c^k$ is the channel gain between the $k$-th cellular UE and the BS, $g^k_{i, c}$ is the interference channel gain between the $i$-th D2D transmitter and the BS in the $k$-th channel. $\sum_{i=1}^{N}p_{i}^k g_{i, c}^k$ denotes the interference from all of the D2D pairs to the BS in the $k$-th channel.  

The achievable rates of the $i$-th D2D pair and the $k$-th cellular UE are given by
\begin{align}
\label{eq:rateD}
r_i^d &=\sum_{k=1}^K \log_2 \left( 1+\gamma_i^k \right),\\
\label{eq:rateC}
r_k^c&=\log_2 \left( 1+\gamma_c^k \right).
\end{align}
 The total power consumption of the $i$-th D2D pair and the $k$-th cellular UE are given by
 \begin{align}
 \label{eq:powerD}
 p_{i,total}^d &= \sum_{k=1}^K \frac{1}{\eta } p_i^k+2p_{cir},\\
 \label{eq:powerC}
  p_{k, total}^c &= \frac{1}{\eta} p_c^k+p_{cir},
 \end{align}
where $p_{i, total}^d$ is the total power consumption of the $i$-th D2D pair, which is composed of the transmission power over all of the $K$ channels, i.e., $\sum_{k=1}^K \frac{1}{\eta } p_i^k$, and the circuit power of both the D2D transmitter and receiver, i.e., $2p_{cir}$. The circuit power of any UE is assumed as the same and is denoted as $p_{cir}$. $\eta$ is the Power Amplifier (PA) efficiency, i.e., $0 < \eta < 1$. $p_{k, total}^c$ is the total power consumption of the $k$-th cellular UE, which is composed of the transmission power $\frac{1}{\eta} p_c^k$ and the circuit power only at the transmitter side. The power consumption of the BS is not taken into consideration.  

\section{Distributed Interference-Aware Energy-Efficient Resource Allocation}
\label{distributed}

\subsection{Problem Formulation}

In the centralized resource allocation, the optimization of the sum EE is carried out by the BS that requires the complete network knowledge. The computational complexity and signaling overhead increase significantly with the number of UEs. Therefore, in this section, we focus on the more practical distributed resource allocation problem, which is modeled as a noncooperative game. 

In the noncooperative game, each UE is self-interested and wants to maximize its own EE. The strategy set of the $i$-th D2D transmitter is denoted as $\mathbf{p}_i^d=\{p_i^k \mid 0 \leq  \sum_{k=1}^K p_i^k  \leq  p_{i, max}^d, k \in \mathcal{K} \}$, $\forall i \in \mathcal{N}$. The strategy set of the $k$-th 
cellular UE is denoted as $\mathbf{p}_k^c=\{ p_c^k \mid 0 \leq p_c^k \leq p_{k, max}^c  \}$, $\forall k \in \mathcal{K}$. $ p_{i, max}^d$ and $p_{k, max}^c$ are the maximum transmission power constraints for D2D UEs and cellular UEs respectively. The strategy set of the other D2D transmitters in $\mathcal{N} \backslash  \{i\}$ is denoted as $\mathbf{p}_{-i}^d=\{ p_j^k \mid  0 \leq  \sum_{k=1}^K p_j^k  \leq  p_{j, max}^d, k \in \mathcal{K}, j \in \mathcal{N}, j \neq i \}$, $\forall i \in \mathcal{N}$. The strategy set of the other cellular UEs in $\mathcal{K} \backslash \{k\}$ is denoted as $\mathbf{p}_{-k}^c=\{p_c^m \mid 0 \leq p_c^m \leq p_{m, max}^c, m\in \mathcal{K}, m\neq k  \}$, $\forall k \in \mathcal{K}$.

For the $i$-th D2D pair, its EE $U_{i, EE}^d$ depends not only on $\mathbf{p}_i^d$, but also on the strategies taken by other UEs in $\mathcal{S}\backslash \{i\}$, i.e.,  $\mathbf{p}_{-i}^d, \mathbf{p}_k^c, \mathbf{p}_{-k}^c$. $U_{i, EE}^d$ is defined as
\begin{align}
\label{eq:UE_EED}
&U_{i, EE}^d (\mathbf{p}_i^d, \mathbf{p}_{-i}^d, \mathbf{p}_k^c, \mathbf{p}_{-k}^c)\notag\\
&=\frac{r_i^d}{p_{i, total}^d}=\frac{\sum_{k=1}^K \log_2 \left( 1+\frac{p_i^k g_{i}^k}{p_c^k g^k_{c, i}+\sum_{j=1, j\neq i}^{N}p_{j}^k g_{j, i}^k+N_0} \right) }{\sum_{k=1}^K \frac{1}{\eta } p_i^k+2p_{cir}}.
\end{align}
Therefore, the EE maximization problem of the $i$-th D2D pair is formulated as
\begin{align}
\label{eq:Dproblem}
&\max. \hspace{10mm} U_{i, EE}^d (\mathbf{p}_i^d, \mathbf{p}_{-i}^d, \mathbf{p}_k^c, \mathbf{p}_{-k}^c) \notag\\
&\mbox{s.t.} \hspace{15mm} C1, C2.
\end{align}

\begin{align}
C1: &r_i^d \geq R_{i, min}^d, \\
C2: &0 \leq \sum_{k=1}^K p_i^k \leq p_{i, max}^d.
\end{align}
Similarly, the EE of the $k$-th cellular UE $U_{k, EE}^c$ is defined as
\begin{align}
U_{k, EE}^c (\mathbf{p}_i^d, \mathbf{p}_{-i}^d, \mathbf{p}_k^c, \mathbf{p}_{-k}^c)=\frac{r_k^c}{p_{k, total}^c}=\frac{\log_2 \left( 1+\frac{p_c^k g_c^k}{\sum_{i=1}^{N}p_{i}^k g_{i, c}^k+N_0} \right)}{\frac{1}{\eta} p_c^k+p_{cir}}.
\end{align}
The corresponding EE maximization problem is formulated as
\begin{align}
\label{eq:Cproblem}
&\max. \hspace{10mm} U_{k, EE}^c (\mathbf{p}_i^d, \mathbf{p}_{-i}^d, \mathbf{p}_k^c, \mathbf{p}_{-k}^c) \notag\\
&\mbox{s.t.} \hspace{15mm} C3, C4.
\end{align}
\begin{align}
C3: &r_k^c \geq R_{k, min}^c,\\
C4: &0 \leq p_c^k \leq p_{k, max}^c.
\end{align}
The constraints C1 and C3 specify the QoS requirements in terms of minimum transmission rate. C2 and C4 are the non-negative constraints on the power allocation variables.

\subsection{Quality of Service Analysis}

In this paper, we have considered the QoS requirement in terms of transmission rate (or equivalently SINR), which is one of the most important metrics and has been widely used in \cite{D2D_Li, Song_Springer2014, Zhou_WCL2014, Wu_TVT2014, Wu_CL2014}. Other important QoS requirements such as delay, or interference threshold can also be expressed as functions of the transmission rate. In this subsection, we investigate relationships among different QoS requirements. Due to the space limitation, multi-QoS based resource allocation schemes are out of the scope of this paper and will be studied in future works.

If we define $T_{i, min}^d$ as the delay tolerance, and there are a total of $B_{i, min}^d$ bits needed to be transmitted by $T_{i, min}^d$. 
Assuming the channel is static during the optimization period, the new EE maximization problem with the QoS requirement in terms of delay is given by
 \begin{align}
 \label{eq:QoS_delay}
&\max. \hspace{5mm} U_{i, EE}^d (\mathbf{p}_i^d, \mathbf{p}_{-i}^d, \mathbf{p}_k^c, \mathbf{p}_{-k}^c) \\
&\mbox{s.t.} \hspace{5mm} C1^{'}: \frac{B_{i, min}^d}{r_i^d}  \leq T_{i, min}^d, \\
 &\hspace{9mm} C2: 0 \leq \sum_{k=1}^K p_i^k \leq p_{i, max}^d.
\end{align}
By rearranging the constraint $C1^{'}$, we have
\begin{align}
r_i^d T_{i, min}^d \geq B_{i, min}^d \Longrightarrow r_i^d \geq \frac{B_{i, min}^d}{T_{i, min}^d} \stackrel{R_{i, min}^d= \frac{B_{i, min}^d}{T_{i, min}^d}}{\Longrightarrow}   r_i^d  \geq R_{i, min}^d.
\end{align}  
Hence, by defining $R_{i, min}^d= \frac{B_{i, min}^d}{T_{i, min}^d}$, we can show that (\ref{eq:QoS_delay}) is equivalent to (\ref{eq:Dproblem}).

Another important QoS requirement is interference threshold, which is particularly important for ensuring proper operation of cellular UEs. If we define $I_{k, max}^c$ as the maximum tolerable interference for the $k$-th cellular UE, the EE maximization problem with the QoS requirement in terms of interference threshold is given by
\begin{align}
\label{eq:QoS_interference}
&\max. \hspace{5mm} U_{k, EE}^c (\mathbf{p}_i^d, \mathbf{p}_{-i}^d, \mathbf{p}_k^c, \mathbf{p}_{-k}^c) \\
&\mbox{s.t.} \hspace{5mm} C3^{'}: \sum_{i=1}^{N}p_{i}^k g_{i, c}^k \leq I_{k, max}^c,\\
 &\hspace{9mm} C4: 0 \leq p_c^k \leq p_{k, max}^c.
\end{align}
Rearranging (\ref{eq:SINRC}), (\ref{eq:rateC}), the interference part $\sum_{i=1}^{N}p_{i}^k g_{i, c}^k$ can be written as a function of $r_i^d$, which is given by
\begin{align}
\label{eq:interferenceC}
\sum_{i=1}^{N}p_{i}^k g_{i, c}^k=\frac{p_c^k g_c^k}{ 2^{r_k^c}-1}-N_0.
\end{align}
By rearranging the constraint $C3^{'}$, we have
\begin{align}
&\sum_{i=1}^{N}p_{i}^k g_{i, c}^k \leq I_{k, max}^c \stackrel{(\ref{eq:interferenceC})}{\Longrightarrow} \frac{p_c^k g_c^k}{ 2^{r_k^c}-1}-N_0 \leq I_{k, max}^c \notag\\
& \Longrightarrow r_k^c \geq \log_2 \left( 1+\frac{p_c^k g_c^k}{I_{k, max}^c+N_0} \right).
\end{align}
Defining $R_{k, min}^c=\log_2 \left( 1+\frac{p_c^k g_c^k}{I_{k, max}^c+N_0} \right)$, $C3^{'}$ can be rewritten as $r_k^c \geq R_{k, min}^c$, which is exactly the same as $C3$. Hence, for the $k$-th cellular UE, we can show that (\ref{eq:QoS_interference}) is equivalent to (\ref{eq:Cproblem}).

\subsection{The Objective Function Transformation}
\label{transformation}

The objective functions in (\ref{eq:Dproblem}) and (\ref{eq:Cproblem}) are non-convex due to the fractional form. In order to derive a closed-form solution, we transform the fractional objective function to a convex optimization function by using the nonlinear fractional programming developed in \cite{Dinkelbach}. We define the maximum EE of the $i$-th D2D pair as $q^{d*}_i$, which is given by
\begin{equation}
q^{d*}_i=\max. U_{i, EE}^d (\mathbf{p}_i^d, \mathbf{p}_{-i}^d, \mathbf{p}_k^c, \mathbf{p}_{-k}^c)=\frac{r_i^d(\mathbf{p}_i^{d*})}{p_{i, total}^d(\mathbf{p}_i^{d*})}.
\end{equation} 
where $\mathbf{p}_i^{d*}$ is the best response of the $i$-th D2D transmitter given the other UEs' strategies $\mathbf{p}_{-i}^d$, $\mathbf{p}_k^c$, $\mathbf{p}_{-k}^c$. The following theorem can be proved:

\textbf{\emph{Theorem 1:}} The maximum EE $q_i^{d*}$ is achieved if and only if 
\begin{align}
\max. \:\: r_i^d (\mathbf{p}_i^d)-q_i^{d*}p_{i, total}^d(\mathbf{p}_i^d)=r_i^d (\mathbf{p}_i^{d*})-q_i^{d*}p_{i, total}^d(\mathbf{p}_i^{d*})=0.
\end{align}
 \begin{IEEEproof}
The proof of Theorem 1 is given in Appendix \ref{theorem1}.
\end{IEEEproof}

 Theorem 1 shows that the transformed problem with an equivalent objective function in subtractive form is equivalent to the non-convex problem, i.e., they lead to the same optimum solution $\mathbf{p}_i^{d*}$.

Similarly, for the maximum EE of the $k$-th cellular UE $q^{c*}_k$, we will have similar theorem as \textbf{\emph{Theorem 1:}}

\textbf{\emph{Theorem 2:}} The maximum EE $q_k^{c*}$ is achieved if and only if 
\begin{align}
\max. \:\: r_k^c (\mathbf{p}_k^c)-q_k^{c*}p_{k, total}^c(\mathbf{p}_k^c)=r_k^c (\mathbf{p}_k^{c*})-q_k^{c*}p_{k, total}^c(\mathbf{p}_k^{c*})=0.
\end{align}
$\mathbf{p}_k^{c*}$ is the best response of the $k$-th cellular UE given the other UEs' strategies $\mathbf{p}_{-k}^c$, $\mathbf{p}_i^d$, $\mathbf{p}_{-i}^d$. $q_i^{d*}$ and $q_k^{c*}$ are not unique \cite{Dinkelbach}.

 \textbf{\emph{Lemma 1:}} The transformed objective function in subtractive form is a concave function.
  \begin{IEEEproof}
The proof of Lemma 1 is given in Appendix \ref{lemma1}.
\end{IEEEproof}

\textbf{\emph{Lemma 2:}} $  \max_{(\mathbf{p}_i^d)} r_i^d (\mathbf{p}_i^d)-q_i^d p_{i, total}^d(\mathbf{p}_i^d)$ is monotonically decreasing as $q_i^d$ increases. 
  \begin{IEEEproof}
The proof of Lemma 2 is given in Appendix \ref{lemma2}.
\end{IEEEproof} 

\textbf{\emph{Theorem 3:}} $F(q_i^d)=\max_{(\mathbf{p}_i^d)} r_i^d (\mathbf{p}_i^d)-q_i^d p_{i, total}^d(\mathbf{p}_i^d)=0$ has a unique solution $q_i^{d*}$.
  \begin{IEEEproof}
The proof of Theorem 3 is given in Appendix \ref{theorem3}.
\end{IEEEproof}

\textbf{\emph{Lemma 3:}} For any feasible $\mathbf{p}_i^{d}$, $\max_{\big(\mathbf{p}_i^{d}\big)} r_i^d \big(\mathbf{p}_i^{d}\big)-q_i^{d} p_{i, total}^d(\mathbf{p}_i^d ) \geq 0$.
  \begin{IEEEproof}
The proof of Lemma 3 is given in Appendix \ref{lemma3}.
\end{IEEEproof}

\subsection{The Iterative Optimization Algorithm}
\label{algorithm} 
The proposed algorithm is summarized in Algorithm \ref{offline algorithm}. $n$ is the iteration index, $L_{max}$ is the maximum number of iterations, and $\Delta$ is the maximum tolerance. 
At each iteration, for any given $q_i^{d}$ or $q_k^{c}$, the resource allocation strategy for the D2D UE or the cellular UE can be obtained by solving the following transformed optimization problems respectively:
\begin{align}
 \label{eq:transformed problemD}
  &\max . \:\: r_i^d (\mathbf{p}_i^d)-q_i^d p_{i, total}^d(\mathbf{p}_i^d) \nonumber\\
 &\mbox{s.t.} \:\:\: C1, C2.
 \end{align}
 \begin{align}
 \label{eq:transformed problemC}
  &\max . \:\: r_k^c (\mathbf{p}_k^c)-q_k^c p_{k, total}^c (\mathbf{p}_k^c) \nonumber\\
 &\mbox{s.t.} \:\:\: C3, C4.
 \end{align}
 
  Taking the D2D UEs as an example, the Lagrangian associated with the problem (\ref{eq:transformed problemD}) is given by
 \begin{align}
&\mathcal{L}_{EE}(\mathbf{p}_i^d, \alpha_i, \beta_i) =r_i^d (\mathbf{p}_i^d)-q_i^d p_{i, total}^d (\mathbf{p}_i^d)\notag\\
&+\alpha_i \left(r_i^d (\mathbf{p}_i^d)-R_{i, min}^d \right) -\beta_i \left( \sum_{k=1}^K p_i^k-p_{i, max}^d\right),
 \end{align}
 where $\alpha_i$, $\beta_i$ are the Lagrange multipliers associated with the constraints C1 and C2 respectively. Since the transformed problem is in a standard concave form with differentiable objective and constraint functions, the Karush-Kuhn-Tucker (KKT) condition are used to find the optimum solutions. The equivalent dual problem can be decomposed into two subproblems: the maximization problem solves the power allocation problem to find the best strategy and the minimization problem solves the master dual problem to find the corresponding Lagrange multipliers, which is given by 
 \begin{equation}
\label{eq:dual problem}
 \displaystyle \min_{\displaystyle (\alpha_i \geq 0, \beta_i \geq 0)}\!\!\!\!. \hspace{5mm} \max_{\displaystyle (\mathbf{p}_i^d)}. \:\:\: \mathcal{L}_{EE}(\mathbf{p}_i^d, \alpha_i, \beta_i) 
\end{equation}

For any given $q_i^d$, the solution is given by
\begin{equation}
\label{eq:waterfilling}
p_i^{k}=\left[ \frac{\eta(1+\alpha_i) \log_2e }{q_i^d+\eta\beta_i }-\frac{p_c^k g^k_{c, i}+\sum_{j=1, j\neq i}^{N}p_{j}^k g_{j, i}^k+N_0}{g_{i}^k}\right]^{+},
\end{equation}
where $[x]^+=\max\{0,x\}$. Equation (\ref{eq:waterfilling}) indicates a water-filling algorithm for transmission power allocation, and the interference from the other UEs decreases the water level. For solving the minimization problem, the Lagrange multipliers can be updated by using the gradient method \cite{improved_step_size, subgradient} as
\begin{align}
\alpha_i (\tau +1)&=\left[ \alpha_i(\tau )-\mu_{i, \alpha} (\tau ) \left( r_i^d(\tau )-R_{i, min}^d \right)    \right]^{+},\\
\beta_i (\tau +1)&=\left[ \beta_i(\tau )+\mu_{i, \beta} (\tau ) \left( \sum_{k=1}^K p_i^k (\tau) - p_{i, max}^d \right)    \right]^{+},
\end{align}
where $\tau$ is the iteration index, $\mu_{i, \alpha}, \mu_{i, \beta}$ are the positive step sizes. The solution of problem (\ref{eq:dual problem}) converges to the optimum solution in (\ref{eq:transformed problemD}) if the step sizes are chosen to satisfy the diminishing step size rules \cite{subgradient}. Since the Lagrange multiplier updating techniques are beyond the scope of this paper, interested readers may refer to \cite{improved_step_size, subgradient} and references therein for details.

Similarly, for any given $q_k^c$, the solution is given by
\begin{equation}
\label{eq:waterfilling_CE}
p_c^k=\left[ \frac{\eta (1+\delta_k) \log_2e }{q_k^c+\eta \theta_k  }-\frac{\sum_{i=1}^N p_{i}^k g_{i, c}^k+N_0}{g_{c}^k} \right]^+,
\end{equation}
where $\delta_k, \theta_k$ are the Lagrange multipliers associated with the constraints C3 and C4 respectively.

\begin{algorithm}[t]
\caption{Iterative Resource Allocation Algorithm}
\label{offline algorithm}
\begin{algorithmic}[1]
\STATE $q_i^d \leftarrow 0$, $q_k^c \leftarrow 0$, $L_{max} \leftarrow 10$, $n \leftarrow 1$, $\Delta \leftarrow 10^{-3}$ 
\FOR{$n=1$ to $L_{max}$}
\IF {D2D link}
\STATE solve (\ref{eq:transformed problemD}) for a given $q_i^d$ and obtain the set of strategies $\mathbf{p}_i^d$
\IF{$r_i^d(\mathbf{p}_i^d)-q_i^d p_{i,total}^d (\mathbf{p}_i^d) \leq \Delta$,}
 \STATE $\mathbf{p}_i^{d*}=\mathbf{p}_i^d$, and $\displaystyle q_i^{d*}=\frac{r_i^d(\mathbf{p}_i^{d*})}{p_{i, total}^d(\mathbf{p}_i^{d*})}$ 
\STATE \textbf{break}
\ELSE
\STATE $\displaystyle q_i^d=\frac{r_i^d(\mathbf{p}_i^d)}{p_{i, total}^d(\mathbf{p}_i^d)}$, and $n=n+1$
\ENDIF
\ELSE
\STATE solve (\ref{eq:transformed problemC}) for a given $q_k^c$ and obtain the set of strategies $\mathbf{p}_k^c$
\IF{$r_k^c(\mathbf{p}_k^c)-q_k^c p_{k,total}^c (\mathbf{p}_k^c) \leq \Delta$,}
 \STATE $\mathbf{p}_k^{c*}=\mathbf{p}_c$, and $\displaystyle q_k^{c*}=\frac{r_k^c (\mathbf{p}_k^{c*})}{p_{k, total}^c(\mathbf{p}_k^{c*})}$ 
\STATE \textbf{break}
\ELSE
\STATE $\displaystyle q_k^c=\frac{r_k^c(\mathbf{p}_k^c)}{p_{k, total}^c(\mathbf{p}_k^c)}$, and $n=n+1$
\ENDIF
\ENDIF
\ENDFOR
\end{algorithmic}
\end{algorithm}

 A Nash equilibrium is a set of power allocation strategies that none UE (neither D2D UE nor cellular UE) can unilaterally improve its EE by choosing a different power allocation strategy, i.e., $\forall i \in \mathcal{N}, \forall k \in \mathcal{K}$, 
\begin{align}
 U_{i, EE}^d(\mathbf{p}_i^{d*}, \mathbf{p}_{-i}^{d*}, \mathbf{p}^{c*}_k, \mathbf{p}_{-k}^{c*}) & \geq  U_{i, EE}^d(\mathbf{p}_i^d, \mathbf{p}_{-i}^d, \mathbf{p}_k^c, \mathbf{p}_{-k}^c), \\
 U_{k, EE}^c(\mathbf{p}_i^{d*}, \mathbf{p}_{-i}^{d*}, \mathbf{p}^{c*}_k, \mathbf{p}_{-k}^{c*})  & \geq  U_{k, EE}^c(\mathbf{p}_i^{d}, \mathbf{p}_{-i}^d, \mathbf{p}^c_k, \mathbf{p}_{-k}^c).
 \end{align}
 
  \textbf{\emph{Theorem 4:}} 
 A Nash equilibrium exists in the noncooperaive game. Furthermore, the strategy set $\{ \mathbf{p}_i^{d*}, \mathbf{p}^{c*}_k \mid  i \in \mathcal{N},  k \in \mathcal{K}\}$ obtained by using Algorithm \ref{offline algorithm} is the Nash equilibrium.
   \begin{IEEEproof}
The proof of Theorem 4 is given in Appendix \ref{theorem4}.
\end{IEEEproof}

\textbf{\emph{Theorem 5:}} The proposed iterative optimization algorithm converges to the optimum EE.
   \begin{IEEEproof}
The proof of Theorem 5 is given in Appendix \ref{theorem5}.
\end{IEEEproof}
 
 \subsection{Complexity Analysis}
\label{complexity_analysis}

The proposed iterative optimization algorithm is based on the nonlinear fractional programming developed in \cite{Dinkelbach}. The iterative algorithm solves the convex problem of (\ref{eq:transformed problemD}) (or (\ref{eq:transformed problemC})) at each iteration, and produces an increasing sequence of $q_i^d$ (or $q_k^c$) values which are proved to converge to the optimum EE (Theorem 5) at a superlinear convergence rate \cite{Dinkelbach_superlinear}. Taking the $i$-th D2D pair as an example, in each iteration, (\ref{eq:transformed problemD}) is solved by using the Lagrange dual decomposition. The algorithmic complexity of this method is dominated by the calculations given by (\ref{eq:waterfilling}), which leads to a total complexity $\mathcal{O} (I_{i, dual}^d I_{i, loop}^{d}  K)$ when $K$ is large, where $I_{i, dual}^d$ is the number of iterations required for reaching convergence, i.e., $I_{i, dual} \leq L_{max}$, and $I_{i, loop}^{d}$ is the required number of iterations for solving the dual problem. 

 In particular, the dual problem (\ref{eq:dual problem}) is decomposed into two subproblems: the inner maximization problem solves the the power allocation problem to find the best strategy and the outer minimization problem solves the master dual problem to find the corresponding Lagrange multipliers. In the inner maximization problem, a total of $I_{i, dual}^d  I_{i, loop}^{d}K (N+3)$ real additions, $I_{i, dual}^d I_{i, loop}^{d}K(N+5)$ real multiplications, and $I_{i, dual}^d I_{i, loop}^{d}K$ real comparisons are required. In the outer minimization problems, a total of $I_{i, dual}^d I_{i, loop}^{d}(K+3)$ real additions, $2 I_{i, dual}^d I_{i, loop}^{d}$ real multiplications, and $2 I_{i, dual}^d I_{i, loop}^{d}$ real comparisons are required. In conclusion, a total of $I_{i, dual}^d I_{i, loop}^{d}(KN+4K+3)$ real additions, $I_{i, dual}^d I_{i, loop}^{d}(KN+5K+2)$ real multiplications, and $I_{i, dual}^d I_{i, loop}^{d}(K+2)$ real comparisons are required for the $i$-th D2D pair.

\subsection{Distributed Implementation}

In the formulated EE maximization problem, the best response of the $i$-th D2D transmitter $\mathbf{p}_i^d$ depends on the strategies of all other UEs, i.e., $\mathbf{p}_{-i}^d, \mathbf{p}_k^c, \mathbf{p}_{-k}^c$. In order to obtain this knowledge, each UE has to broadcast its transmission strategy to other UEs. However, we observe that the sufficient information of $\mathbf{p}_{-i}^d, \mathbf{p}_k^c, \mathbf{p}_{-k}^c$ are contained in the form of interference, i.e., $p_c^k g^k_{c, i}$ and $\sum_{j=1, j\neq i}^{N} p_{j}^k g_{j, i}^k$. In this way, each D2D pair has only to estimate the interference on all available channels to determine the power optimization rather than knowing the specific strategies of other UEs. For the $k$-th cellular UE, the BS estimates the interference from D2D pairs on the $k$-th channel and then feeds back this information to the cellular UE. If UEs update their strategies sequentially, player strategies will eventually converge to a Nash equilibrium, which is proved to exist in Theorem 4. The D2D peer discovery techniques and the design of strategy updating mechanism are out of the scope of this paper and will be discussed in future works.

\subsection{Efficiency Analysis}

  One useful solution for evaluating the efficiency of a Nash equilibrium is the price of anarchy.  The price of anarchy is defined as the ratio of the maximum social welfare, i.e., sum EE of the overall network, achieved by a centralized resource allocation scheme to the EE achieved at the worst-case equilibrium \cite{Gametheory_Han}.

The EE of the overall network is a function of the power allocation strategies, which is given by
\begin{align}
 \label{eq:U_EE}
 U_{EE}(\mathcal{P}_d, \mathcal{P}_c)=\sum_{i=1}^N \frac{r_i^d}{ p_{i,total}^d }+\sum_{k=1}^K \frac{r_k^c}{p_{k, total}^c}, 
\end{align}
where $\mathcal{P}_d$ and $\mathcal{P}_c$ are the sets of power allocation strategies for D2D UEs and cellular UEs respectively, i.e., $\mathcal{P}_d=\{p_i^k \mid 0 \leq  \sum_{k=1}^K p_i^k  \leq  p_{i, max}^d,    i \in \mathcal{N}, k \in \mathcal{K}\}$, $\mathcal{P}_c=\{p_k^c \mid 0 \leq p_c^k \leq p_{k, max}^c, k \in \mathcal{K}\}$.  
This definition of (\ref{eq:U_EE}) is not based on the ratio of sum network throughput to sum network power consumption as in \cite{Feiran_2012, Wu_TVT2014}, because transmission power and achievable rates can not be shared among UEs \cite{Miao_TWC2011}. 

Taking (\ref{eq:SINRD}), (\ref{eq:SINRC}), (\ref{eq:rateD}), (\ref{eq:rateC}), (\ref{eq:powerD}), and (\ref{eq:powerC}) into (\ref{eq:U_EE}), the EE of the overall network is rewritten as
\begin{align}
 \label{eq:U_EE_full}
 U_{EE}(\mathcal{P}_d, \mathcal{P}_c)& =\sum_{i=1}^N \frac{\sum_{k=1}^K \log_2 \left( 1+\frac{p_i^k g_{i}^k}{p_c^k g^k_{c, i}+\sum_{j=1, j\neq i}^{N}p_{j}^k g_{j, i}^k+N_0} \right) }{\sum_{k=1}^K \frac{1}{\eta } p_i^k+2p_{cir}}\notag\\
&+\sum_{k=1}^K \frac{\log_2 \left( 1+\frac{p_c^k g_c^k}{\sum_{i=1}^{N}p_{i}^k g_{i, c}^k+N_0} \right)}{\frac{1}{\eta} p_c^k+p_{cir}}.
\end{align}
The $U_{EE}$ defined in (\ref{eq:U_EE_full}) is not a concave function for $p_i^k, p_c^k$ ($p_i^k \in \mathcal{P}_d, p_c^k \in \mathcal{P}_c)$, and it is intractable to find the global maximum EE of the overall network. However, we can get some insights about energy-efficient power allocation design by considering some special cases. The price of anarchy for the general case is analyzed through computer simulations.

%\textcolor[cmyk]{0, 1, 1, 0}{\subsubsection{The Transmission Power Dominated Case}
%\label{tran_case}
%The transmission power dominated case represents that $p_i^k, p_c^k >> p_{cir}$, $\forall i \in \mathcal{N}, \forall k \in \mathcal{K}$. Since $U_{EE}$ is strictly decreasing with $p_i^k, p_c^k$ (it can be proved that $\frac{\partial U_{EE}}{\partial p_i^k}<0$ and $\frac{\partial U_{EE}}{\partial p_c^k}<0$), the optimum strategy is to use as little power as possible subject to the QoS constraint. Similarly, it can be proved that either $U_{i, EE}^d$ or $U_{k, EE}^c$ is strictly decreasing with $p_i^k$ or $p_c^k$. Thus, the optimum strategy for the distributed algorithm is the same as that of the centralized algorithm. Therefore, in the transmission power dominated case, the price of anarchy is $1$. }

 \subsubsection{Noise Dominated Case}
\label{noise_case}
The noise dominated case represents that $N_0>>p_c^k g_{c, i}^k+\sum_{j=1, j\neq i}^N p_j^k g_{j, i}^k$, $N_0>>\sum_{i=1}^N p_i^k g_{i,c}^k$, $\forall i \in \mathcal{N}, \forall k \in \mathcal{K}$. Thus, the EE maximization problem in the noise dominated case is decomposed into independent $N+K$ subproblems, which is equivalent to the solution of the distributed algorithm. Therefore, in the noise dominated case, the price of anarchy is $1$. 

\subsubsection{Cellular UE Dominated Case}
\label{C_case}
The cellular UE dominated case arises in scenarios where a cellular UE is far from the BS but close to the D2D pair, and the transmission power of cellular UEs is much stronger than the transmission power of the D2D transmitter, i.e., $p_c^k>>p_i^k, p_c^k g_{c, i}^k >> p_i^k g_i^k $, $\forall i \in \mathcal{N}, \forall k \in \mathcal{K}$. The D2D UEs are forced to stop transmission due to the severe interference caused by cellular UEs, which solely occupy all of the available channels. Thus, the EE maximization problem can be decomposed into independent $K$ subproblems, which is equivalent to the solution of the distributed algorithm. Therefore, in the cellular UE dominated case, the price of anarchy is $1$.

\section{Distributed Interference-Aware Spectral-Efficient Resource Allocation}
\label{distributed_SE}

In this section, for the purpose of comparison, we derive the distributed interference-aware spectral-efficient resource allocation by employing the noncooperative game model developed in Section \ref{distributed}. Each UE is self-interested and wants to maximize its own SE rather than EE, and the power consumption is completely ignored in the optimization process. For the $i$-th D2D pair, its SE utility function $U_{i, SE}^d$ depends not only on $\mathbf{p}_i^d$, but also on the strategies taken by other UEs in $\mathcal{S}\backslash \{i\}$, i.e., $\mathbf{p}_{-i}^d$, $\mathbf{p}_{k}^c$, $\mathbf{p}_{-k}^c$. $U_{i, SE}^d$ is defined as
\begin{align}
\label{eq:UE_SED}
&U_{i, SE}^d (\mathbf{p}_i^d, \mathbf{p}_{-i}^d, \mathbf{p}_k^c, \mathbf{p}_{-k}^c) \notag\\
&=r_i^d=\sum_{k=1}^K \log_2 \left( 1+\frac{p_i^k g_{i}^k}{p_c^k g^k_{c, i}+\sum_{j=1, j\neq i}^{N}p_{j}^k g_{j, i}^k+N_0} \right).
\end{align}

Therefore, the SE maximization problem of the $i$-th D2D pair is formulated as
\begin{align}
\label{eq:Dproblem_SE}
&\max. \hspace{10mm} U_{i, SE}^d (\mathbf{p}_i^d, \mathbf{p}_{-i}^d, \mathbf{p}_k^c, \mathbf{p}_{-k}^c) \notag\\
&\mbox{s.t.} \hspace{15mm} C1, C2.
\end{align}
Similarly, the SE of the $k$-th cellular UE $U_{k, SE}^c$ is defined as
\begin{align}
U_{k, SE}^c (\mathbf{p}_i^d, \mathbf{p}_{-i}^d, \mathbf{p}_k^c, \mathbf{p}_{-k}^c)=r_k^c=\log_2 \left( 1+\frac{p_c^k g_c^k}{\sum_{i=1}^{N}p_{i}^k g_{i, c}^k+N_0} \right).
\end{align}
The corresponding SE maximization problem is formulated as
\begin{align}
\label{eq:Cproblem_SE}
&\max. \hspace{10mm} U_{k, SE}^c (\mathbf{p}_i^d, \mathbf{p}_{-i}^d, \mathbf{p}_k^c, \mathbf{p}_{-k}^c) \notag\\
&\mbox{s.t.} \hspace{15mm} C3, C4.
\end{align}

It is noted that the objective functions in (\ref{eq:Dproblem_SE}) and (\ref{eq:Cproblem_SE}) are concave and closed-form solution can be derived by exploiting the properties of convex optimization. Taking the D2D UEs as an example, given the other UEs' strategies $\mathbf{p}_{-i}^d$, $\mathbf{p}_k^c$, $\mathbf{p}_{-k}^c$, the Lagrangian associated with the problem (\ref{eq:Dproblem_SE}) is given by
 \begin{align}
&\mathcal{L}_{SE}(\mathbf{p}_i^d, \alpha_i, \beta_i )\notag\\
&=r_i^d (\mathbf{p}_i^d)+\alpha_i \left( r_i^d(\mathbf{p}_i^d)-R_{i, min}^d \right) - \beta_i \left(\sum_{k=1}^K p_i^k-p_{i, max}^d \right),
 \end{align}
 where $\alpha_i$, $\beta_i$ are the Lagrange multipliers associated with the constraints C1 and C2 respectively. The equivalent Lagrange dual problem:
 \begin{equation}
\label{eq:dual_SE_D}
 \displaystyle \min_{\displaystyle (\alpha_i \geq 0, \beta_i \geq 0)}\!\!\!\!. \hspace{5mm} \max_{\displaystyle (\mathbf{p}_i^d)}. \:\:\: \mathcal{L}_{SE}(\mathbf{p}_i^d, \alpha_i, \beta_i). 
\end{equation}
The dual problem in (\ref{eq:dual_SE_D}) can be decomposed into two subproblems: the maximization problem solves the power allocation problem to find the best strategy and the minimization problem solves the master dual problem to find the corresponding Lagrange multipliers. For any given $\alpha_i, \beta_i$, the solution is given by
\begin{equation}
\label{eq:waterfilling_SE}
p_i^{k*}=\left [ \frac{(1+\alpha_i)\log2e}{\beta_i}-\frac{p_c^{k*} g^k_{c, i}+\sum_{j=1, j\neq i}^{N}p_{j}^k g_{j, i}^k+N_0}{g_i^k}  \right]^{+}.
\end{equation}
Equation (\ref{eq:waterfilling_SE}) indicates a water-filling algorithm for transmission power allocation, and the interference from the other UEs decreases the water level. The Lagrange multipliers can be updated by using the gradient method introduced in Section \ref{distributed}.

Similarly, the optimum solution of $p_c^{k*}$ is given by
\begin{equation}
\label{eq:waterfilling_SE_C}
p_c^{k*}=\left[ \frac{ (1+\delta_k) \log_2e }{\theta_k }-\frac{\sum_{i=1}^N p_{i}^{k*} g_{i, c}^k+N_0}{g_{c}^k} \right]^+,
\end{equation}
where $\delta_k, \theta_k$ are the Lagrange multipliers associated with the constraints C3 and C4 respectively.

 A Nash equilibrium is a set of power allocation strategies that none UE (neither D2D UE nor cellular UE) can unilaterally improve its SE by choosing a different power allocation strategy, i.e., $\forall i \in \mathcal{N}, \forall k \in \mathcal{K}$, 
\begin{align}
 U_{i, SE}^d(\mathbf{p}_i^{d*}, \mathbf{p}_{-i}^{d*}, \mathbf{p}^{c*}_k, \mathbf{p}_{-k}^{c*}) & \geq  U_{i, SE}^d(\mathbf{p}_i^d, \mathbf{p}_{-i}^d, \mathbf{p}_k^c, \mathbf{p}_{-k}^c), \\
 U_{k, SE}^c(\mathbf{p}_i^{d*}, \mathbf{p}_{-i}^{d*}, \mathbf{p}^{c*}_k, \mathbf{p}_{-k}^{c*})  & \geq  U_{k, SE}^c(\mathbf{p}_i^{d}, \mathbf{p}_{-i}^d, \mathbf{p}^c_k, \mathbf{p}_{-k}^c).
 \end{align}

\textbf{\emph{Theorem 6:}} 
 A Nash equilibrium exists in the noncooperaive game. Furthermore, the strategy set $\{ \mathbf{p}_i^{d*}, \mathbf{p}^{c*}_k \mid  i \in \mathcal{N},  k \in \mathcal{K}\}$ obtained by (\ref{eq:waterfilling_SE}), (\ref{eq:waterfilling_SE_C}) is the Nash equilibrium.
\begin{IEEEproof}
The proof of Theorem 6 is given in Appendix \ref{theorem6}.
\end{IEEEproof}

\section{Energy Efficiency and Spectral Efficiency Tradeoff}
\label{tradeoff}

In this section, we investigate the tradeoff between EE and SE. For the $i$-th D2D pair, the EE gap between the energy-efficient algorithm and the spectral-efficient algorithm, which are derived in Section \ref{distributed} and Section \ref{distributed_SE} respectively, is defined as
\begin{align}
\label{eq:EE_D_gap}
G_{i, EE}^d&=U_{i, EE}^{d*}-\frac{U_{i, SE}^{d*}}{(p_{i, total}^d)_{SE}}\notag\\
&=\frac{\sum_{k=1}^K \log_2 \left( 1+\frac{p_{i, EE}^{k*} g_{i}^k}{p_{c, EE}^{k*} g^k_{c, i}+\sum_{j=1, j\neq i}^{N}p_{j, EE}^{k*} g_{j, i}^k+N_0} \right) }{\sum_{k=1}^K \frac{1}{\eta } p_{i, EE}^{k*}+2p_{cir}}\notag\\
&-\frac{\sum_{k=1}^K \log_2 \left( 1+\frac{p_{i, SE}^{k*} g_{i}^k}{p_{c, SE}^{k*} g^k_{c, i}+\sum_{j=1, j\neq i}^{N}p_{j, SE}^{k*} g_{j, i}^k+N_0} \right) }{\sum_{k=1}^K \frac{1}{\eta } p_{i, SE}^{k*}+2p_{cir}},
\end{align}  
where $U_{i, EE}^{d*}$ and $U_{i, SE}^{d*}$ are the maximum EE and SE which are obtained by solving the problems in (\ref{eq:Dproblem}) and (\ref{eq:Dproblem_SE}) respectively. $p_{i, EE}^{k*}$ and $p_{c, EE}^{k*}$ are the optimum energy-efficient power allocation solution given by Algorithm 1 (using (\ref{eq:waterfilling}) and (\ref{eq:waterfilling_CE}) respectively). $p_{i, SE}^{k*}$ and $p_{c, SE}^{k*}$ are the optimum spectral-efficient power allocation solution given by (\ref{eq:waterfilling_SE}) and (\ref{eq:waterfilling_SE_C}) respectively. The SE gap between the spectral-efficient algorithm and the energy-efficient algorithm is defined as 
\begin{align}
\label{eq:SE_D_gap}
G_{i, SE}^d&=U_{i, SE}^{d*}-(p_{i, total}^d)_{EE}U_{i, EE}^{d*}\notag\\
&=\sum_{k=1}^K \log_2 \left( 1+\frac{p_{i, SE}^{k*} g_{i}^k}{p_{c, SE}^{k*} g^k_{c, i}+\sum_{j=1, j\neq i}^{N}p_{j, SE}^{k*} g_{j, i}^k+N_0} \right) \notag\\
&-\sum_{k=1}^K \log_2 \left( 1+\frac{p_{i, EE}^{k*} g_{i}^k}{p_{c, EE}^{k*} g^k_{c, i}+\sum_{j=1, j\neq i}^{N}p_{j, EE}^{k*} g_{j, i}^k+N_0} \right) .
\end{align}
Similarly, for the $k$-th cellular UE, the EE and SE gaps between the energy-efficient and the spectral-efficient algorithms are given by
\begin{align}
\label{eq:EE_C_gap}
G_{k, EE}^c&=U_{k, EE}^{c*}-\frac{U_{k, EE}^{c*}}{(p_{k, total}^c)_{SE}}\notag\\
&=\frac{\log_2 \left( 1+\frac{p_{c, EE}^{k*} g_c^k}{\sum_{i=1}^{N}p_{i, EE}^{k*} g_{i, c}^k+N_0} \right)}{\frac{1}{\eta} p_{c, EE}^{k*}+p_{cir}}
-\frac{\log_2 \left( 1+\frac{p_{c, SE}^{k*} g_c^k}{\sum_{i=1}^{N}p_{i, SE}^{k*} g_{i, c}^k+N_0} \right)}{\frac{1}{\eta} p_{c, SE}^{k*}+p_{cir}},
\end{align}
\begin{align}
\label{eq:SE_C_gap}
G_{k, SE}^c&=U_{k, SE}^{c*}-(p_{k, total}^c)_{EE}U_{k, EE}^{c*}\notag\\
&=\log_2 \left( 1+\frac{p_{c, SE}^{k*} g_c^k}{\sum_{i=1}^{N}p_{i, SE}^{k*} g_{i, c}^k+N_0} \right)\notag\\
&-\log_2 \left( 1+\frac{p_{c, EE}^{k*} g_c^k}{\sum_{i=1}^{N}p_{i, EE}^{k*} g_{i, c}^k+N_0} \right),
\end{align}
where $U_{k, EE}^{c*}$ and $U_{k, SE}^{c*}$ are the maximum EE and SE which are obtained by solving (\ref{eq:Cproblem}) and (\ref{eq:Cproblem_SE}) respectively.

Although the EE and SE gaps can be calculated by using (\ref{eq:EE_D_gap}), (\ref{eq:SE_D_gap}), (\ref{eq:EE_C_gap}), (\ref{eq:SE_C_gap}), the numerical results depends on the specific channel realization in each simulation and a large number of simulations are required to obtain the average result. In order to facilitate analysis and get some insights, we consider a special case that all the signal channels  have the same power gain $g$, and all the interference channels have the same power gain $\hat{g}$. The interference level of the overall network is defined as $I=\frac{\hat{g}}{g}$.  The EE and SE gaps defined in (\ref{eq:EE_D_gap}), (\ref{eq:SE_D_gap}), (\ref{eq:EE_C_gap}), (\ref{eq:SE_C_gap}) can be rewritten as
\begin{align}
\label{eq:EE_D_gap_ideal}
G_{i, EE}^d&=\frac{K\log_2 \left( 1+\frac{p_{i, EE}^{k*}} {p_{c, EE}^{k*} I+(N-1) p_{i, EE}^{k*} I+\frac{N_0}{g}} \right) }{\frac{K}{\eta } p_{i, EE}^{k*}+2p_{cir}}\notag\\
&-\frac{K \log_2 \left( 1+\frac{p_{i, SE}^{k*} }{p_{c, SE}^{k*} I+Np_{j, SE}^{k*} I+\frac{N_0}{g}} \right) }{\frac{K}{\eta } p_{i, SE}^{k*}+2p_{cir}},
\end{align}  
\begin{align}
\label{eq:SE_D_gap_ideal}
G_{i, SE}^d&=K \log_2 \left( 1+\frac{p_{i, SE}^{k*} }{p_{c, SE}^{k*} I+(N-1) p_{i, SE}^{k*} I+\frac{N_0}{g}} \right) \notag\\
&-K\log_2 \left( 1+\frac{p_{i, EE}^{k*}}{p_{c, EE}^{k*} I+Np_{i, EE}^{k*}I+\frac{N_0}{g}} \right) ,
\end{align}
\begin{align}
\label{eq:EE_C_gap_ideal}
G_{k, EE}^c=\frac{\log_2 \left( 1+\frac{p_{c, EE}^{k*} }{Np_{i, EE}^{k*}I+\frac{N_0}{g}} \right)}{\frac{1}{\eta} p_{c, EE}^{k*}+p_{cir}}
-\frac{\log_2 \left( 1+\frac{p_{c, SE}^{k*}}{Np_{i, SE}^{k*} I+\frac{N_0}{g}} \right)}{\frac{1}{\eta} p_{c, SE}^{k*}+p_{cir}},
\end{align}
\begin{align}
\label{eq:SE_C_gap_ideal}
G_{k, SE}^c&=\log_2 \left( 1+\frac{p_{c, SE}^{k*} }{Np_{i, SE}^{k*} I+\frac{N_0}{g}} \right)\notag\\
&-\log_2 \left( 1+\frac{p_{c, EE}^{k*} }{Np_{i, EE}^{k*}I+\frac{N_0}{g}} \right).
\end{align}
The relationships among the EE and SE tradeoff, the EE and SE gap, and the interference level are analyzed through simulations by using the above the equations derived above.

\section{Simulation Results}
\label{Simulation Results}

\begin{figure}[t]
\begin{center}
\scalebox{0.63} 
{\includegraphics{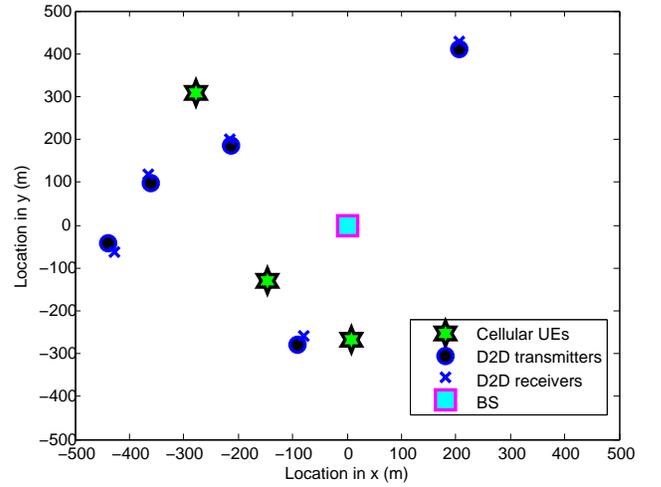}}
\end{center}
\caption{The locations of D2D UEs and cellular UEs generated in one simulation ($N=5$, $K=3$, the cell radius is $500$ m, and maximum D2D distance is $25$ m ). A total of $1000$ simulations are performed.}
\label{scenario}
\end{figure}

In this section, the proposed algorithm is verified through computer simulations. The values of simulation parameters are inspired by \cite{Feiran_WCNC2013, Feiran_2012, Song_JSAC} , and are summarized in Table \ref{simulation_parameters}. We compare the proposed EE maximization algorithm (labeled as ``energy-efficient") with the SE maximization algorithm (labeled as ``spectral-efficient" ), and the random power allocation algorithm (labeled as ``random"). The results are averaged through a total number of $1000$ simulations and  normalized by the maximum value. For each simulation, the locations of the cellular UEs and D2D UEs are generated randomly within a cell with a radius of $500$ m. Fig. \ref{scenario} shows the locations of D2D UEs and cellular UEs generated in one simulation. The maximum distance between any two D2D UEs that form a D2D pair is $25$ m. The channel gain between the transmitter $i$ and the receiver $j$ is calculated as $d_{i, j}^{-2} |h_{i, j}|^2$ \cite{Feiran_WCNC2013, Feiran_2012, Feiran_ICC2013}, where $d_{i, j}$ is the distance between the transmitter $i$ and the receiver $j$, $h_{i, j}$ is the complex Gaussian channel coefficient that satisfies $h_{i, j} \sim \mathcal{CN} (0, 1)$.

\begin{table}[t]
\caption{Simulation Parameters.}
\label{simulation_parameters}
\begin{center}
\begin{tabular}{|l|l|}
\hline
\textbf{Parameter}&\textbf{Value}\\
\hline
Cell radius & 500 m\\
\hline
Maximum D2D transmission distance & 25 m\\
\hline
Maximum transmission power $\displaystyle p_{i,max}^d, \displaystyle p_{k, max}^c$ & 200 mW (23 dBm)\\
\hline
Constant circuit power $p_{cir}$ &10 mW (10 dBm)\\
\hline
Thermal noise power $N_0$ & $\displaystyle 10^{-7}$ W\\
\hline
Number of D2D pairs $N$ & 5\\
\hline
Number of cellular UEs $K$ & 3\\
\hline
PA efficiency $\eta $ & 35\%\\
\hline
QoS of cellular UEs $\displaystyle R_{k, min}^c$ & 0.1 bit/s/Hz\\
\hline
QoS of D2D UEs $\displaystyle R_{i, min}^d$ & 0.5 bit/s/Hz\\
\hline
\end{tabular}
\end{center}
\end{table}

\begin{figure}[t]
\begin{center}
\scalebox{0.63} 
{\includegraphics{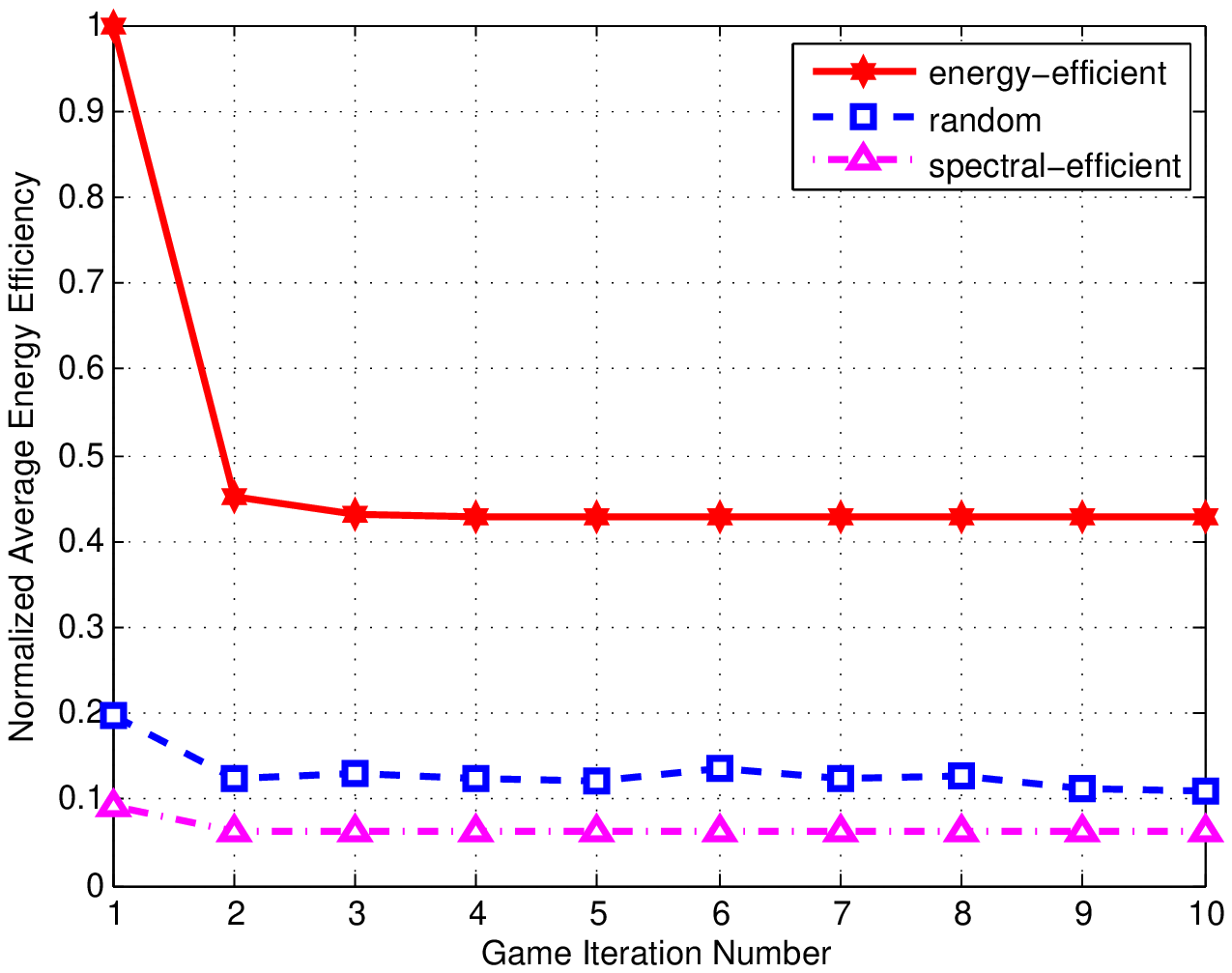}}
\end{center}
\caption{The normalized average energy efficiency of D2D links corresponding to the number of game iterations ($N=5$, $K=3$, $p_{i, max}^d=p_{k, max}^c=200$ mW, $R_{k, min}^c=0.1$ bit/s/Hz, $R_{i, min}^d=0.5$ bit/s/Hz, $1000$ simulations).}
\label{EE_D2D}
\end{figure}

Fig. \ref{EE_D2D} shows the normalized average EE of D2D links corresponding to the number of game iterations. The normalized average EE of the proposed energy-efficient algorithm converge to $0.429$, while the random algorithm converge to $0.124$ and the spectral-efficient algorithm converge to $0.064$. It is clear that the proposed energy-efficient algorithm significantly outperforms the spectral-efficient algorithm and the random algorithm in terms of EE in an interference-limited environment. The spectral-efficient algorithm has the worst EE performance among the three because power consumption is completely ignored in the optimization process. 

\begin{figure}[t]
\begin{center}
\scalebox{0.63} 
{\includegraphics{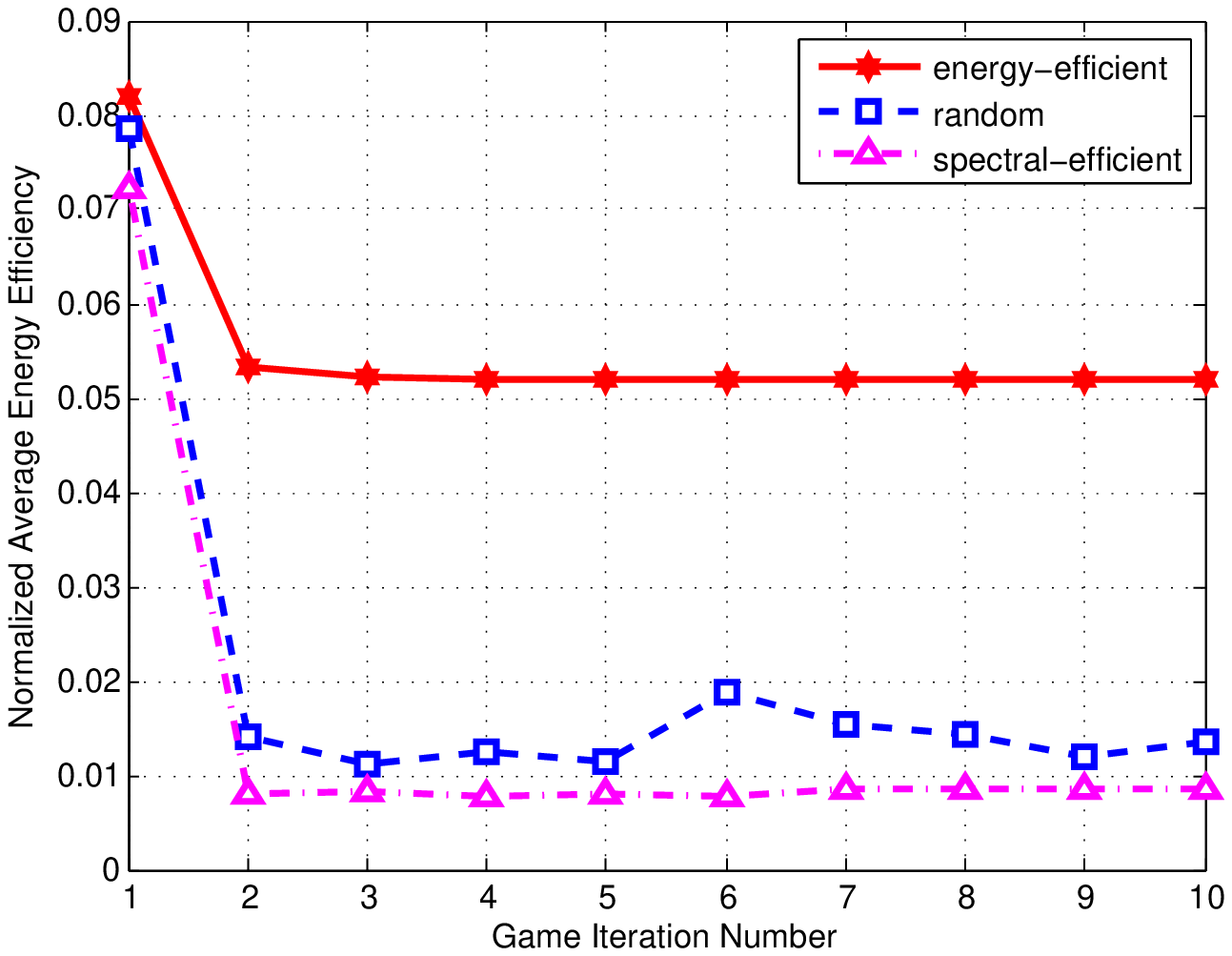}}
\end{center}
\caption{The normalized average energy efficiency of cellular links corresponding to the number of game iterations ($N=5$, $K=3$, $p_{i, max}^d=p_{k, max}^c=200$ mW, $R_{k, min}^c=0.1$ bit/s/Hz, $R_{i, min}^d=0.5$ bit/s/Hz, $1000$ simulations).}
\label{EE_CE}
\end{figure}

\begin{figure}[t]
\begin{center}
\scalebox{0.63} 
{\includegraphics{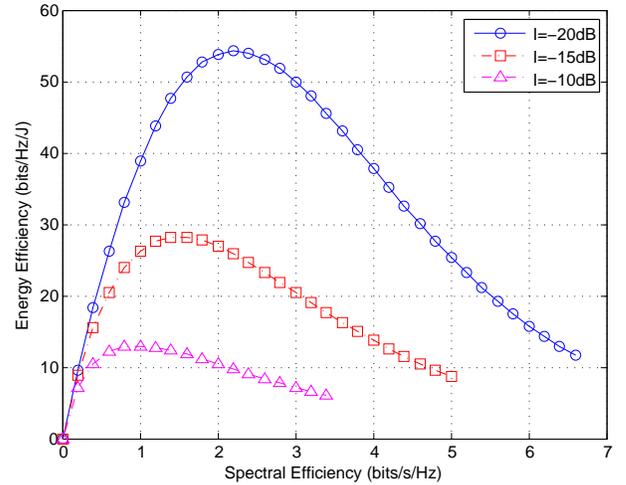}}
\end{center}
\caption{The energy efficiency and spectral efficiency tradeoff for cellular UEs corresponding to three interference levels $I=-20, -15, -10$ dB, ($g=1, N=1, K=1$, $p_{i, max}^d=p_{k, max}^c=200$ mW).}
\label{EE_SE_tradeoff}
\end{figure}

Fig. \ref{EE_CE} shows the normalized average EE of cellular links corresponding to the number of game iterations. The simulation results demonstrate that the proposed algorithm achieves the best performance again. Comparing Fig. \ref{EE_CE} with Fig. \ref{EE_D2D}, we find that the D2D links can achieve a much better EE than the cellular links due to the proximity gain and the channel  reuse gain. The proximity gain is achieved by shorter transmission distance, while the channel reuse gain is achieved by proper interference management. The proposed energy-efficient algorithm and the conventional SE algorithm converge to the equilibrium within $3 \sim 4$ game iterations, while the random algorithm fluctuates around the equilibrium since that the transmission power strategy is randomly selected. Although power consumption is also ignored in the random algorithm, the random algorithm performs better than the spectral-efficient algorithm. The reason is explained in Fig. \ref{EE_SE_tradeoff}.

\begin{figure}[t]
\begin{center}
\scalebox{0.63} 
{\includegraphics{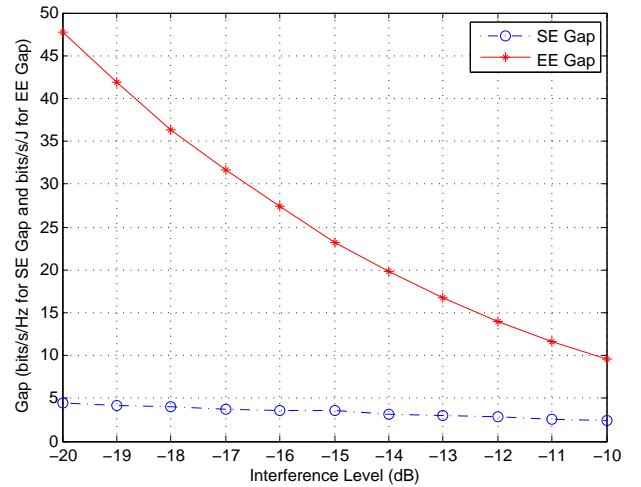}}
\end{center}
\caption{The energy efficiency and spectral efficiency gaps of the cellular UE with regards to the interference level $I$ ($g=1, N=1, K=1$, $p_{i, max}^d=p_{k, max}^c=200$ mW).}
\label{GAP2}
\end{figure}

\begin{figure}[t]
\begin{center}
\scalebox{0.63} 
{\includegraphics{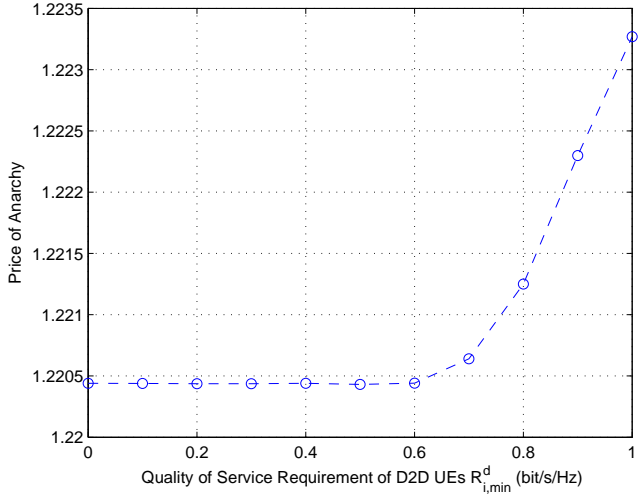}}
\end{center}
\caption{The price of anarchy with regards to the QoS requirement ($R_{k, min}^c=R_{i, min}^d/5, N=1, K=1$, $p_{i, max}^d=p_{k, max}^c=200$ mW, 1000 simulations).}
\label{POA}
\end{figure}

Fig. \ref{EE_SE_tradeoff} shows the tradeoff between EE and SE for the cellular UE under different interference scenarios, i.e., $\mathbf{I= -20, -15, -10}$ dB. We consider the special case discussed in Section \ref{tradeoff}. The SE of the cellular UE is increased from $0$ bits/s/Hz to $7$ bits/s/Hz with a step of $0.2$, and the corresponding transmission power $p_c^k$ is calculated by using (\ref{eq:SINRC}) and (\ref{eq:rateC}). We assume that the D2D transmitter is selfish and always use the maximum transmission power. For each step of SE, the corresponding EE is obtained through simulations. In this special case, the channel gains are fixed and the maximum achievable SE is limited by the transmission power constraint. For example, when $I=-15$ dB, the maximum achievable $U_{k, SE}^c$ is only $5.0728$ bits/s/Hz. Simulation results are infeasible when $U_{k, SE}^c \geq 6$ bits/s/Hz.

For the case of $\mathbf{I=-20}$ dB, the maximum achievable SE and EE subject to the transmission power constraint are $6.6$ bits/s/Hz and $54.26$ bits/s/J respectively. In comparison, for the case of $\mathbf{I=-15}$ dB, the maximum achievable SE and EE are $5$ bits/s/Hz and $28.21$ bits/s/J respectively. By increasing the interference level from $\mathbf{-20}$ dB to $\mathbf{-15}$ dB, the maximum achievable SE and EE are reduced by nearly  $24\%$ and $48\%$ respectively. 
We conclude that as interference level increases, the EE decreases more rapidly than the SE. 
Furthermore, if we further increase the transmission power, the EE degrades severely while the SE only improves slightly. For example, when $\mathbf{I=-20}$ dB, if we increase the SE from $2.2$ bits/s/Hz to $4$ bits/s/Hz, the corresponding EE is reduced from $54.26$ bits/s/J to $37.83$ bits/s/J. As a result, the SE is only increased by $1.8$ bits/s/Hz, but the EE is reduced by $16.43$ bits/Hz/J. Hence, increasing transmission power beyond the power for optimum EE brings little SE improvement but significant EE loss. However, in the severe interference case, i.e., $\mathbf{I=-10}$ dB, the EE loss is not so large due to the fact that the maximum achievable EE is limited by the interference. 

 Fig. \ref{EE_SE_tradeoff} also explains why the random algorithm performs better than the spectral-efficient algorithm. Taking the case of $I=-15$ dB as an example, the spectral-efficient algorithm always select the point with maximum EE, (i.e., SE$=5$ bits/s/Hz, EE$=8.618$ bits/Hz/J). Among the $26$ points on the curve, there is only one point (SE=$0$ bits/s/Hz, EE=$0$ bits/Hz/J), whose EE performance is worse. In other words, if we randomly select one point out of these $26$ points, the probability of having a higher EE than the spectral-efficient algorithm is $24/26 \approx 92\%$. For the case of $I=-20$ dB and $I=-10$ dB, the probability is approximately $91\%$ and $89\%$ respectively. Therefore, this shows that the random algorithm performs better than the spectral-efficient algorithm.

Fig. \ref{GAP2} shows the EE and SE gaps of the cellular UE (defined in (\ref{eq:EE_C_gap_ideal}) and (\ref{eq:SE_C_gap_ideal}) respectively) with regards to the interference level $I$. From Fig. \ref{GAP2}, it is clear that both the EE and SE gaps ($G_{i, EE}^c$ and $G_{i, SE}^c$) decrease as the interference level $I$ increasing. In particular, the EE gap decreases much more rapidly than the SE gap, which verifies again that in an interference-limited environment, increasing transmission power beyond the power for optimum EE brings little SE improvement but significant EE loss. Therefore, the proposed energy-efficient algorithm can bring significant EE improvement subject to little SE loss.

 Fig. \ref{POA} shows the price of anarchy with regards to the QoS requirements $R_{i,min}^d$ and $R_{k,min}^c$. $R_{i,min}^d$ is increased from $0$ to $1$ bit/s/Hz with a step of 0.1, and $R_{k,min}^c=R_{i,min}^d/5$. The exhaustive optimum sum EE is used for comparison. The simulation result indicates that the proposed distributed algorithm provides high system efficiency (the price of anarchy is close to 1). Moreover, the price of anarchy is stable below 1.23, and only increases slightly as QoS requirement increases. The reason is that as QoS requirement increases, both cellular UEs and D2D UEs become aggressive, which leads to the performance degradation of the distributed algorithm.

\section{Conclusion}
\label{Conclusion}
In this paper, we proposed a distributed interference-aware energy-efficient resource allocation algorithm for D2D communications by exploiting the properties of the nonlinear fractional programming. Simulation results have demonstrated that the proposed energy-efficient algorithm significantly outperforms the spectral-efficient algorithm in terms of EE for both cellular and D2D links. We have analyzed the tradeoff between EE and SE and derived closed-form expressions for EE and SE gaps. Through simulation results we found that in an interference-limited environment, increasing transmission power beyond the power for optimum EE brings little SE improvement but significant EE loss. Therefore, the proposed energy-efficient algorithm can bring significant EE improvement subject to little SE loss.

% or


\appendices
\section{Proof of the Theorem 1}
\label{theorem1}
The proof of the Theorem 1 is similar to the proof of the Theorem (page 494 in \cite{Dinkelbach}). Firstly, we prove the necessity proof. For any feasible strategy set $\mathbf{p}_i^d$, $\forall i \in \mathcal{N}$, we have 
\begin{align}
\label{eq:sb}
q_i^{d*}=\frac{r_i^d(\mathbf{p}_i^{d*})}{p_{i, total}^d(\mathbf{p}_i^{d*})} \geq \frac{r_i^d(\mathbf{p}_i^d)}{p_{i, total}^d(\mathbf{p}_i^d)}.
\end{align}
By rearranging (\ref{eq:sb}), we obtain
\begin{align}
r_i^d(\mathbf{p}_i^{d*})-q_i^{d*}p_{i, total}^d(\mathbf{p}_i^{d*})&=0,\\
r_i^d(\mathbf{p}_i^{d})-q_i^{d*}p_{i, total}^d(\mathbf{p}_i^{d}) &\leq 0.
\end{align}
Hence, the maximum value of $r_i^d(\mathbf{p}_i^{d})-q_i^{d*}p_{i, total}^d(\mathbf{p}_i^{d})$ is $0$, and can only be achieved by $\mathbf{p}_i^{d*}$, which is obtained by solving the EE maximization problem defined in (\ref{eq:Dproblem}). This completes the necessity proof.

Now we turn to the sufficiency proof. Assume that $\mathbf{\tilde{p}}_i^{d}$ is the optimal solution which satisfies that
\begin{align}
\label{eq:assumption_theorem1}
r_i^d(\mathbf{p}_i^{d})-q_i^{d*}p_{i, total}^d(\mathbf{p}_i^{d}) &\leq r_i^d(\mathbf{\tilde{p}}_i^{d})-q_i^{d*}p_{i, total}^d(\mathbf{\tilde{p}}_i^{d}) =0.
\end{align}
By rearranging (\ref{eq:assumption_theorem1}), we have
\begin{align}
q_i^{d*}=\frac{r_i^d(\mathbf{\tilde{p}}_i^{d})}{p_{i, total}^d(\mathbf{\tilde{p}}_i^{d})} \geq \frac{r_i^d(\mathbf{p}_i^d)}{p_{i, total}^d(\mathbf{p}_i^d)}.
\end{align}
Hence, $\mathbf{\tilde{p}}_i^{d}$ is also the solution of the EE maximization problem defined in (\ref{eq:Dproblem}), i.e., $\mathbf{\tilde{p}}_i^{d}=\mathbf{p}_i^{d*}$. This completes the sufficiency proof.

\section{Proof of the Lemma 1}
\label{lemma1}

 Taking $r_i^d (\mathbf{p}_i^d)-q_i^d p_{i, total}^d(\mathbf{p}_i^d)$ as an example, which is the transformed objective function in subtractive form corresponding to the $i$-th D2D pair. The first part $r_i^d (\mathbf{p}_i^d)$ can be rewritten as 
\begin{align}
r_i^d (\mathbf{p}_i^d)=\sum_{k=1}^K \log_2 \left( 1+\frac{p_i^k g_{i}^k}{p_c^k g^k_{c, i}+\sum_{j=1, j\neq i}^{N}p_{j}^k g_{j, i}^k+N_0} \right),
\end{align}
which is the sum of $K$ concave functions. The second part $-q_i^d p_{i, total}^d(\mathbf{p}_i^d)$ is given by
\begin{align}
-q_i^d p_{i, total}^d(\mathbf{p}_i^d)=-q_i^d \left( \sum_{k=1}^K \frac{1}{\eta } p_i^k+2p_{cir} \right),
\end{align}
 which is the sum of $K$ affine functions. Since the sum of a concave function and an affine function is also concave, this completes the proof of Lemma 1.
 
 \section{Proof of the Lemma 2}
\label{lemma2}

Define $q_i^{d*}<q_i^{d*'}$, and define $\mathbf{p}_i^{d*}$ and $\mathbf{p}_i^{d*'}$ as the corresponding optimum solutions respectively. We have
\begin{align}
& \max_{(\mathbf{p}_i^d)} r_i^d (\mathbf{p}_i^d)-q_i^{d*} p_{i, total}^d(\mathbf{p}_i^d) =  r_i^d (\mathbf{p}_i^{d*})-q_i^{d*} p_{i, total}^d(\mathbf{p}_i^{d*}) \notag\\
&>  r_i^d (\mathbf{p}_i^{d*'})-q_i^{d*} p_{i, total}^d(\mathbf{p}_i^{d*'})>  r_i^d (\mathbf{p}_i^{d*'})-q_i^{d*'} p_{i, total}^d(\mathbf{p}_i^{d*'})\notag\\
& = \max_{(\mathbf{p}_i^d)} r_i^d (\mathbf{p}_i^d)-q_i^{d*'} p_{i, total}^d(\mathbf{p}_i^d). 
\end{align}

\section{Proof of the Theorem 3}
\label{theorem3}

We have the following fact: $\lim_{q_i^d \to -\infty} F(q_i^d) = + \infty$, and $\lim_{q_i^d \to +\infty} F(q_i^d) = - \infty$. Since $F(q_i^d)$ is monotonically decreasing as $q_i^d$ increases and continuous for $q_i^d$, $F(q_i^d)=0$ has a unique solution $q_i^{d*}$. Otherwise, if we assume that $\hat{q}_i^{d*} \neq q_i^{d*}$, and $F(\hat{q}_i^{d*})=0$, according to Lemma 1 and Lemma 2, we must  either have $F(q_i^{d*})=0>F(\hat{q}_i^{d*})$ (if $\hat{q}_i^{d*}>q_i^{d*}$), or $F(q_i^{d*})=0<F(\hat{q}_i^{d*})$ if ($\hat{q}_i^{d*}<q_i^{d*}$). This contradicts with the assumption that $\hat{q}_i^{d*} \neq q_i^{d*}$, and $F(\hat{q}_i^{d*})=0$.

 \section{Proof of the Lemma 3}
\label{lemma3}

Define an feasible solution $\mathbf{\hat{p}}_i^{d}$ such that $q_i^d=\frac{r_i^d (\mathbf{\hat{p}}_i^{d})}{p_{i, total}^d (\mathbf{\hat{p}}_i^{d}) }$, we have
\begin{align}
 \max_{\big(\mathbf{p}_i^{d}\big)} r_i^d \big(\mathbf{p}_i^{d}\big)-q_i^{d} p_{i, total}^d(\mathbf{p}_i^d )
\geq  r_i^d \big(\mathbf{\hat{p}}_i^{d}\big)-q_i^{d} p_{i, total}^d(\mathbf{\hat{p}}_i^d )=0.
\end{align}

\section{Proof of the Theorem 4}
\label{theorem4}

According to \cite{game_theory_1994}, a Nash equilibrium exists if the utility function is continuous and quasiconcave, and the set of strategies is a nonempty compact convex subset of a Euclidean space. Taking the EE objection function defined in (\ref{eq:UE_EED}) as an example, the numerator $r_i^d$ defined in (\ref{eq:rateD}) is a concave function of $p_i^k$, $\forall i \in \mathcal{N}, k \in \mathcal{K}$. The denominator defined in (\ref{eq:powerD}) is an affine function of $p_i^k$. Therefore, $U_{i, EE}^d$ is quasiconcave (Problem 4.7 in \cite{convex_optimization}). The set of the strategies $\mathbf{p}_i^d=\{p_i^k \mid 0 \leq  \sum_{k=1}^K p_i^k  \leq  p_{i, max}^d, k \in \mathcal{K} \}$, $\forall i \in \mathcal{N}$, is a nonempty compact convex subset of the Euclidean space $\mathbb{R}^K$. Similarly, it is easily proved that the above conditions also hold for the cellular UE. Therefore, a Nash equilibrium exists in the noncooperaive game.
 
  If the strategy set $\mathbf{p}_i^{d*}$ obtained by using Algorithm \ref{offline algorithm} is not the Nash equilibrium, the $i$-th D2D transmitter can choose the Nash equilibrium $\mathbf{\hat{p}}_i^{d}$ ($\mathbf{\hat{p}}_i^{d} \neq \mathbf{p}_i^{d*}$) to obtain the maximum EE $q_i^{d*}$. However, by Theorem 1, $q_i^{d*}$ can only be achieved by choosing $\mathbf{p}_i^{d*}$. Then, we must have $\mathbf{\hat{p}}_i^{d} = \mathbf{p}_i^{d*}$, which contradicts with the assumption. Therefore, $\mathbf{p}_i^{d*}$ is part of the Nash equilibrium. A similar proof holds for $\mathbf{p}^{c*}_k$. It is proved that the set $\{ \mathbf{p}_i^{d*}, \mathbf{p}^{c*}_k \mid  i \in \mathcal{N},  k \in \mathcal{K}\}$ obtained by using Algorithm \ref{offline algorithm} is the Nash equilibrium.
  
  \section{Proof of the Theorem 5}
\label{theorem5}

Firstly, we prove that the EE for the $i$-th D2D pair $q_i^d$ increases in each iteration. We denote that $\mathbf{\hat{p}}_i^{d}(n)$ as the optimum resource allocation policies in the $n$-th iteration, and $q_i^{d*}$ as the optimum EE. We denote $q_i^d (n)$ and $q_i^d (n+1)$ as the EE in the $n$-th iteration and $(n+1)$-th iteration respectively, and we assume that $q_i^d (n) \neq q_i^{d*}$, and $q_i^d (n+1) \neq q_i^{d*}$. $q_i^d (n+1)$ is updated in the $n$-th iteration of the proposed Algorithm 1 as $q_{n+1}=\frac{r_i^d \big(\mathbf{\hat{p}}_i^{d}(n)\big)}{p_{i, total}^d \big(\mathbf{\hat{p}}_i^{d}(n)\big) }$. We have
 \begin{align}
  & \max_{\big(\mathbf{p}_i^{d}(n)\big)} r_i^d \big(\mathbf{p}_i^{d}(n)\big)-q_i^{d}(n) p_{i, total}^d(\mathbf{p}_i^d (n))\notag\\
&= r_i^d \big(\mathbf{\hat{p}}_i^{d}(n)\big)-q_i^d (n) p_{i, total}^d \big( \mathbf{\hat{p}}_i^{d}(n) \big) \notag\\
 &=q_i^d (n+1)p_{i, total}^d \big(\mathbf{\hat{p}}_i^{d}(n)\big)-q_i^d (n) p_{i, total}^d \big( \mathbf{\hat{p}}_i^{d}(n)\big) \notag\\
&=p_{i, total}^d \big(\mathbf{\hat{p}}_i^{d}(n)\big) \big( q_i^d (n+1)-q_i^d (n) \big) \stackrel{\mathrm{Theorem 1, Lemma 2, lemma 3}}{>}0
 \notag\\
 &\stackrel{\mathrm{p_{i, total}^d \big(\mathbf{\hat{p}}_i^{d}(n)\big)>0}}{\Longrightarrow }  q_i^d (n+1)>q_i^d (n)   
 \end{align}
 
 Secondly, by combining $q_i^d (n+1)>q_i^d (n)$, Lemma 2, and Lemma 3, we can prove that 
\begin{align}
& \max_{\big(\mathbf{p}_i^{d}\big)} r_i^d \big(\mathbf{p}_i^{d}\big)-q_i^{d}(n) p_{i, total}^d(\mathbf{p}_i^d )\notag\\
&> \max_{\big(\mathbf{p}_i^{d}\big)} r_i^d \big(\mathbf{p}_i^{d}\big)-q_i^{d}(n+1) p_{i, total}^d(\mathbf{p}_i^d ) \notag\\
&> \max_{\big(\mathbf{p}_i^{d}\big)} r_i^d \big(\mathbf{p}_i^{d}\big)-q_i^{d*} p_{i, total}^d(\mathbf{p}_i^d )=r_i^d \big(\mathbf{p}_i^{d*}\big)-q_i^{d*} p_{i, total}^d(\mathbf{p}_i^{d*} )=0.
\end{align}
Therefore, $q_i^d (n) $ is increased in each iteration and will eventually approaches $q_i^{d*}$ as long as $L_{max}$ is large enough, and $\max_{\big(\mathbf{p}_i^{d}\big)} r_i^d \big(\mathbf{p}_i^{d}\big)-q_i^{d} p_{i, total}^d(\mathbf{p}_i^d )$ will approach zero and satisfy the optimality conditions proved in Theorem 1.

\section{Proof of the Theorem 6}
\label{theorem6}

According to \cite{game_theory_1994}, a Nash equilibrium exists if the utility function is continuous and quasiconcave, and the set of strategies is a nonempty compact convex subset of a Euclidean space. Taking the SE objection function defined in (\ref{eq:UE_SED}) as an example, $r_i^d$ defined in (\ref{eq:rateD}) is a concave function of $p_i^k$, $\forall i \in \mathcal{N}, k \in \mathcal{K}$. Therefore, $U_{i, EE}^d$ is quasiconcave since any concave function is quasiconcave \cite{convex_optimization}. The set of the strategies $\mathbf{p}_i^d=\{p_i^k \mid 0 \leq  \sum_{k=1}^K p_i^k  \leq  p_{i, max}^d, k \in \mathcal{K} \}$, $\forall i \in \mathcal{N}$, is a nonempty compact convex subset of the Euclidean space $\mathbb{R}^K$. Similarly, it is easily proved that the above conditions also hold for the cellular UE. Therefore, a Nash equilibrium exists in the noncooperaive game.

If the strategy set $\mathbf{p}_i^{d*}$ obtained by (\ref{eq:waterfilling_SE}) is not the Nash equilibrium, the $i$-th D2D transmitter can choose the Nash equilibrium $\mathbf{\hat{p}}_i^{d}$ ($\mathbf{\hat{p}}_i^{d} \neq \mathbf{p}_i^{d*}$) to obtain the maximum SE defined in (\ref{eq:Dproblem_SE}). Hence, $\mathbf{\hat{p}}_i^{d}$ is also the solution of the SE maximization problem defined in (\ref{eq:Dproblem_SE}), i.e., $\mathbf{\hat{p}}_i^{d}=\mathbf{p}_i^{d*}$. This completes the proof.

\section*{Acknowledgment}
This work was partially supported by National Science Foundation of China (NSFC) under Grant Number 61203100, 61450110085, the Open Research Project of the State Key Laboratory of Industrial Control Technology, Zhejiang University, China (No. ICT1407), JSPS KAKENHI Grant Number 25880002, 26730056 and JSPS A3 Foresight Program, and Fundamental Research Funds for the Central Universities under Grant Number 13MS19, 14MS11, China Mobile Communication Co. Ltd.  Research Institute (CMRI), and China Electric Power Research Institute (CEPRI) of State Grid Corporation of China (SGCC).

\ifCLASSOPTIONcaptionsoff
  \newpage
\fi



%\bibliographystyle{IEEEtran}

%\begin{biography}[{\includegraphics[width=1in,height=1.25in,clip,keepaspectratio]{mshell}}]{Michael Shell}
\bibliographystyle{IEEEtran}
\bibliography{IEEE_gc_2014}

%\end{IEEEbiography}


%\newpage



%\vfill


\end{document}